\documentstyle[preprint,aps]{revtex}
\tightenlines

\newcommand{\be}{\begin{equation}}
\newcommand{\ee}{\end{equation}}
\newcommand{\bea}{\begin{eqnarray}}
\newcommand{\eea}{\end{eqnarray}}

\begin{document}

\begin{titlepage}
\begin{center}
\vskip .2in
\hfill
\vbox{
    \halign{#\hfil         \cr
           hep-th/9712073 \cr
           SU-ITP-97-50 \cr
           SLAC-PUB-7712\cr
           December 1997    \cr
           }  
      }   
\vskip 0.5cm
{\large \bf SPECIAL CONFORMAL SYMMETRY OF WORLDVOLUME ACTIONS}\\
\vskip 1.0cm
R. Kallosh\footnote{email address: kallosh@physics.stanford.edu}, J. Kumar
\footnote{email address: jkumar@leland.stanford.edu}

{\em Department of Physics, Stanford University, Stanford, California 94305 USA
\\}
\vskip 0.5cm
 A. Rajaraman\footnote{email address: arvindra@dormouse.stanford.edu}

{\em Stanford Linear Accelerator Center, Stanford University, Stanford, CA
94309 \\}

\vskip 1cm

\end{center}
\begin{abstract}
Kappa-symmetric worldvolume  actions of the  D3-, M5- and M2-branes can
be coupled consistently to their   near horizon bosonic geometry background.
We study the gauge-fixed action in the approximation in which
only the  transverse radial direction of the brane is allowed to fluctuate. The
generalized special conformal symmetry of these self-interacting actions is
established.
This opens up a possibility to find out if the full superconformal symmetry of
the free actions of these branes survives in the presence of coupling defined
by the size of the anti-deSitter throat.

\vskip 0.5cm
 \end{abstract}
\end{titlepage}
\newpage

Recently the relation between  superconformal symmetry of the worldvolume
actions and near horizon geometry of  branes was studied \cite{M5tens}
\cite{Maldacena}.

In \cite{M5tens} the kappa-symmetric M5-brane action \cite{5b} was gauge-fixed
in the Killing gauge \cite{RenataQuant} in the flat 11-dimensional target
space. This
resulted in a self-interacting non-linear action of (0,2) tensor multiplet in
d=6. The full non-linear action has 32 worldvolume supersymmetries, half of
which are of Volkov-Akulov type.
The quadratic part of this action was shown to have full superconformal
symmetry. However the interaction terms were shown to break the superconformal
symmetry. It was suggested that the quantization of the M5-brane action in the
curved background may change the situation and some generalization of the rigid
superconformal symmetry may exist for the interacting theory. The indication of
this was found via the study of special conformal transformations using the
properties of the $adS$ geometry. It was found that one must use a particular field
dependent modification of the standard special conformal symmetry (see eq. (6.13)
in \cite{M5tens}):

\begin{equation}
 \Lambda^\alpha_K =  \Lambda^\alpha _- + {(6\mu)^2 \over \phi^2(x)}
\Lambda^\alpha_+
\label{adesitter}\end{equation}
Here $ \Lambda^\alpha _-$ is a global parameter of special conformal symmetry,
$\phi(x) $ is related to a radial coordinate of the brane, which fluctuates on
the
worldvolume\footnote{We renamed here the field $w(x)$ of \cite{M5tens} into
$\phi(x)$  to avoid confusion with conformal weight $w$.} and $\mu$ defines the
size of the $adS$ throat.

On the other hand, an important observation was made in \cite{Maldacena}. If one
has
a scaling invariant action of the type
\begin{equation}
S = b \int d^{4}x U^{4} \left [\sqrt{ (1-  \tilde R^4 \partial _\alpha U
\partial
^\alpha U/U^{4}) }- 1\right]~,
\label{juan}\end{equation}
one can verify the invariance of this action under the generalized special
conformal symmetry  with additional field dependent terms of the form shown in
eq. (\ref{adesitter})
\begin{eqnarray}
\delta x^\alpha  &=& \epsilon^\beta x_\beta  x^\alpha -
\epsilon^\alpha ( x^2 -{ \tilde R^4 \over U^2 })/2 \\
\delta U & \equiv& U'(x') -U(x)   = - \epsilon^\alpha x_\alpha U ~,
\end{eqnarray}
where $\epsilon^\alpha$ is an infinitesimal parameter.

But the relation of the action (\ref{juan}) to
the gauge-fixed kappa symmetric action of the D3 brane, which has a unique
supersymmetric generalization, remained unclear.  Additionally, one might
wish to check if the
actions of other branes which have a near-horizon $AdS_{p+2}\times S^{d-p-2}$
geometry, such as the M-5 brane or M-2 brane, are also conformal.

One can establish generically that when only radial excitations are allowed, the
bosonic part of such an action \cite{kappaactions} in
the background of the near horizon geometry is

\begin{equation}
S= - C \int d^{p+1} x  \left( { U\over \tilde R}\right)^{{p+1\over w}}
 \left[ \sqrt{ -Det \left ( \eta_{\alpha \beta} + \tilde R^{p+1\over w}
 {\partial_\alpha U  \partial_\beta U\over U^{p+1\over w}}     \right )} -1
\right]
\qquad
w={p-1\over 2}
\label{action}\end{equation}
with $\alpha, \beta = 0,1, ... ,p$.  The conformal weight of $U$
is $w$. The radial direction on the brane is the only non-Killing direction
since the geometry is spherically symmetric and depends only on radial
coordinate.

We can simplify the expression under the square root:
\be
-Det \left ( \eta_{\alpha \beta} +X_{\alpha \beta}    \right ) =  Det \left (
\delta_{\alpha}{}^{ \gamma} +X_{\alpha}{}^{ \gamma}    \right )  (-Det (\eta_{
\gamma \beta})) = Det \left ( \delta_{\alpha}{}^{ \gamma} +X_{\alpha}{}^{
\gamma}    \right ) \ee
where
\be
X_{\alpha \beta} \equiv \tilde R^{p+1\over w} {\partial_\alpha
U  \partial_\beta U\over U^{p+1\over w}}
\ee
We then use the identity
\be
Det (\eta_{\alpha}{ }^{\beta} + X_{\alpha}{ }^{\beta} )= exp(Tr{  }log(
\eta_{\alpha}{ }^{\beta} + X_{\alpha}{ }^{\beta} )) = exp{ }Tr(
\sum_{n=1} {{(-1)^{n+1}\over n}X^{n}})
\ee
and we note that
\begin{eqnarray}
Tr(X^{n}) = \left({\tilde R^{p+1\over w} \over U^{p+1\over w}}
\right)^{n}
(\partial_{\alpha_1 } U)(\partial^{\alpha_2 }
U)(\partial_{\alpha 2 } U)(\partial^{\alpha_3 } U) ... (\partial_{\alpha_{n} }
U)
(\partial^{\alpha_1 } U) = (Tr{ }X)^{n}
\end{eqnarray}

Thus we find that the logarithm and trace operations commute, and

\be
Det (\eta_{\alpha}{ }^{\beta} + X_{\alpha}{ }^{\beta} )= exp({ }log({ }
1 + Tr X_{\alpha}{ }^{\beta}))) =
1 + Tr X_{\alpha}{ }^{\beta}
\ee

We thus conclude that the truncated action may be written as

\be
S = -C \int d^{p+1}x \left( { U\over \tilde R}\right)^{{p+1\over w}} \left[
\sqrt{1+\tilde R^{p+1\over w}
{(\partial_{\alpha}
U)(\partial^{\alpha} U)\over U^{p+1\over w} }  }-1\right]
\label{truncated}\ee

This action is invariant under the following special
conformal transformation:

\begin{eqnarray} \label{special}
\delta x^\alpha  &=& \epsilon^\beta x_\beta  x^\alpha -
\epsilon^\alpha ( x^2 +{ w^2 \tilde R^{p+1\over w} \over U^{2\over w}})/2 \nonumber\\
\delta U & \equiv& U'(x') -U(x)   = -w \epsilon^\alpha x_\alpha U ~,
\end{eqnarray}

One can see this simply by using the substitution

\be
U=\phi^{w}
\ee

where $\phi $ is now a weight 1 field.  We then see that

\be
{(\partial_{\alpha} U)(\partial^{\alpha} U)\over U^{2+ {2\over w}}}
= w^2 \phi^{2w-2} {(\partial_{\alpha} \phi)(\partial^{\alpha} \phi)\over
\phi^{2w+2}} = w^2 {(\partial_{\alpha} \phi)(\partial^{\alpha} \phi)\over
\phi^{4}}
\ee
\begin{eqnarray}
S = -C^{\prime } \int d^{p+1}x \phi^{p+1} \left[ \sqrt{1+ \mu^4
{(\partial_{\alpha}
\phi)(\partial^{\alpha} \phi)\over \phi^4 }  }
-1\right] \ , \qquad
\mu^4 = w^2 \tilde R^{p+1\over w}
\end{eqnarray}

The proof of conformal invariance of this truncated action under the above
transformation (up to a different choice of a sign \footnote{Our choice of a
sign between two terms in square root is defined by the original action
(\ref{action}) with the square root of the determinant.}) was shown in
\cite{Maldacena}.

Thus we have found that part of the bosonic actions of the gauge-fixed
kappa-symmetric D3- and M5- and M2-branes has a generalized special conformal
symmetry of the type shown in eq. (\ref{adesitter}) as predicted by the study
of the anti-deSitter geometry and as found in \cite{Maldacena} for a closely
related theory (\ref{juan}). This opens up an exciting possibility that there
exist non-linear interacting superconformal
worlvolume field theories, waiting to be discovered. There are two
possibilities to proceed. The first one, indicated in \cite{Maldacena} is to
generalize the action (\ref{action}) to the full superconformal symmetry. The
second one is to use the available kappa-symmetric actions of the branes and to
work out the  quantization of kappa-symmetry in $AdS_{p+2}\times S^{d-p-2}$
backgrounds following the simpler case developed in \cite{RenataQuant} for  the
flat background. This procedure will automatically supply us with the
supersymmetric version of the action (\ref{action}) which  either will be
superconformally symmetric or not. An indication that it might be
superconformally symmetric is given by the fact that the supersymmetric
actions in quadratic approximation have a full superconformal symmetry. This
corresponds to the approximation where the size of the anti-deSitter throat can
be sent to zero. In our case here, at $\tilde R \rightarrow 0$ the truncated
actions (\ref{truncated}) of the D3- and M2- and M5-brane tend to

\be
S = C \int d^{p+1}x \left(  -{1\over 2}(\partial_{\alpha} U) ^2+ {1\over 8}
\tilde R^{p+1\over w} {[(\partial_{\alpha} U)^2]^2 \over U^{p+1\over w} }
-{1\over 16} (\tilde R^{( p+1)\over w})^2 {[(\partial_{\alpha} U)^2 ]^3 \over
U^{2(p+1)\over w} } + \dots \right)
\ee

which at $\tilde R = 0$ has special conformal symmetry under the transformation
(\ref{special}) with $\tilde R = 0$. In this approximation we know that all 3
type of free supersymmetric actions do have a superconformal symmetry. For the
M2-brane there is a superconformal scalar multiplet action in d=3, for the
D3-brane there is a superconformal vector multiplet action in d=4. For the
M5-brane (0,2) tensor multiplet  the superconformal symmetry of the free action
 was  established recently \cite{M5tens}. The size of the anti-deSitter throat
$\tilde R$ plays the role of the coupling constant for these non-gravitational
theories.  It remains to be seen if the superconformal symmetry of free actions
can be generalized
in the presence of coupling as it already happened for the special conformal
part of the symmetry, presented in eq. (\ref{special}) for the approximation of
the theory with only radial excitations.

 \vskip 0.5 cm
This work  is supported by the NSF grant PHY-9219345. The work of J. K. was
also supported by the Department of Defense, NDSEG Program.  The work of A. R. was
supported in part by the Department of Energy under contract No. DE-AC03-76SF00515.

\end{document}